\documentclass[referee]{aa} 
\input psfig.sty
\begin{document}
\title{Radio Observations and Spectrum of the SNR G127.1+0.5 and its Central Source 0125+628}
\author{Denis Leahy \inst{1}
\and
Wenwu Tian \inst{1,2}}
      
\authorrunning{D.A. Leahy and W.W. Tian}
\offprints{W. W. Tian}
\institute{Department of Physics \& Astronomy, University of Calgary, Calgary, Alberta T2N 1N4, Canada \\
\and
National Astronomical Observatories, CAS, Beijing 100012, China 
}
 
\date{Received Dec. 7, 2005; accepted Jan. 20, 2006} 
\abstract{
We present new images of the Supernova Remnant (SNR) G127.1+0.5 (R5), based on the 408 MHz and 1420 MHz continuum emission and the HI-line emission data of the Canadian Galactic Plane Survey (CGPS).
The radio spectrum of the central compact source (0125+628) is analyzed in
the range 178 MHz - 8.7 GHz, indicating a flat spectrum with synchrotron
self-absorption below 800 MHz.
The SNR's flux density at 408 MHz is
17.1$\pm$1.7 Jy and at 1420 MHz is 10.0$\pm$0.8 Jy, corrected for flux densities from compact sources within the SNR.
The SNR's integrated flux density based spectral index
(S$_{\nu}$$\propto$$\nu$$^{-\alpha}$) is 0.43$\pm$0.10.
The respective T-T plot spectral index (derived from the relative size of brightness
temperature variations between two frequencies, see text for details) is 0.46$\pm$0.01.  There is no evidence at 1$\sigma$ for spatial variations in spectral index within
G127.1+0.5. In particular, we compared the northern shell, southern shell and central diffuse region.  HI observations show structures associated with the SNR in the radial velocity range of -12 to -16
km$/$s, suggesting G127.1+0.5's distance is 1.15 kpc. The estimated Sedov age
is 2 - 3 $\times$$10^{4}$ yr.

}
\titlerunning{Radio Spectrum of the SNR G127.1+0.5}
\maketitle 

\section{Introduction}
We report new CGPS radio observations 
of the  bilateral symmetry shell-type SNR G127.1+0.5 (R5). 
G127.1+0.5 is considered as a younger remnant (18000 yr, Milne 1988) for its appearance and brightness. Because compact sources are detected near the center of the SNR, 
there have been many studies at radio, optical and X-ray wavelengths of the SNR and 
the compact sources (Pauls 1977; Pauls et al., 1982; Fu\"rst et al., 1984; Joncas et al., 1989; Xilouris et al., 1993; Kaplan et al. 2004). But its basic physical features are still uncertain, 
such as its distance and age and radio spectrum. 
G127.1+0.5's radio spectral index in frequency range of 0.4 GHz - 5 GHz has been estimated as
between 0.45 and 0.6 previously (F\"urst et al., 1984; Joncas et al., 1989).  
The spectrum of the southern shell is slightly steeper than for the northern shell based on observations with a competitive resolution but lower sensitivity than current observations. 
In this paper, we present the SNR's continuum images at higher sensitivity than previously 
at both 408 MHz and 1420 MHz in order to determine its flux densities and spectrum. 
We search the HI-line emission for detecting possible interactions of the remnant with the surrounding gas and estimating its distance and age. 

\section{Observations and Analysis}
The continuum and HI emission data sets come from the CGPS,
which is described in detail by Taylor et al. (2003).
The data sets are mainly based on observations from the Synthesis Telescope (ST) of the Dominion Radio Astrophysical Observatory (DRAO). The angular resolution of the continuum images is better than 1$^{\prime}$$\times$ 1$^{\prime}$ cosec($\delta$) (HPBW) at 1420 MHz and 3.4$^{\prime}$$\times$3.4$^{\prime}$ cosec($\delta$) at 408 MHz. The synthesized beam for the HI line images is as the same as for the continuum  and the radial velocity resolution is 1.32 km$/$s. DRAO ST observations are not sensitive to structures larger than an angular size scale of about 3.3$^{o}$ at 408 MHz and 56$^{\prime}$  at 1420 MHz. Thus the CGPS includes data from the 408 MHz all-sky survey of Haslam et al (1982), sensitive to structure greater 51$^{\prime}$, and the Effelsberg 1.4 GHz Galactic plane survey of Reich et al. (1990, 1997), sensitive to structure with 9.4$^{\prime}$ for large scale emission (the single-dish data are freely available by http://www.mpifr-bonn.mpg.de/survey.html).  The low-order spacing HI data is from the single-antenna survey of the CGPS area (Higgs $\&$ Tapping 2000) with resolution of 36$^{\prime}$.  See Taylor et al. (2003) for detail of the method of combining the synthesis telescopes and single dish observations. 

We analyze the continuum and HI images of G127.1+0.5 and determine its flux
densities using the DRAO export software package.  
For G127.1+0.5, integrated flux density's errors are found by comparing results for several different choices of background region. We also consider a systematic error of about 5$\%$ contributed by the flux calibration uncertainties at 408 MHz and 1420 MHz. For compact sources, the flux density's errors are taken as the formal Gaussian fit errors. We consider the influence of compact sources within the SNR by employing similar methods to Tian and Leahy (2005a), although the compact sources within G127.1+0.5 are not bright compared with those for  another nearby SNR G126.2+1.6 (Tian and Leahy, 2005b).
 
\begin{figure*}
\vspace{55mm}
\begin{picture}(200,300)
\put(-40,125){\includegraphics{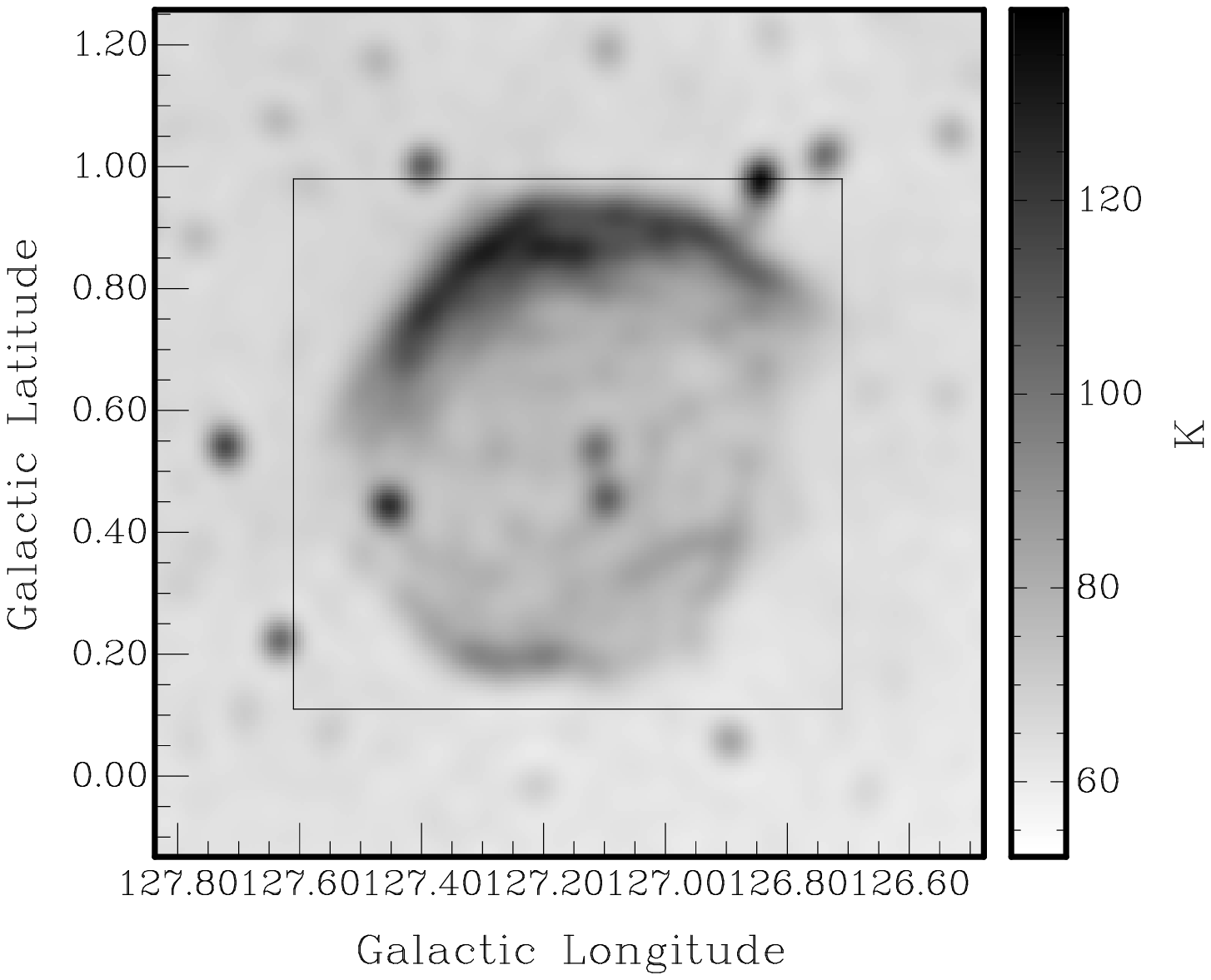}}
\put(215,125){\includegraphics{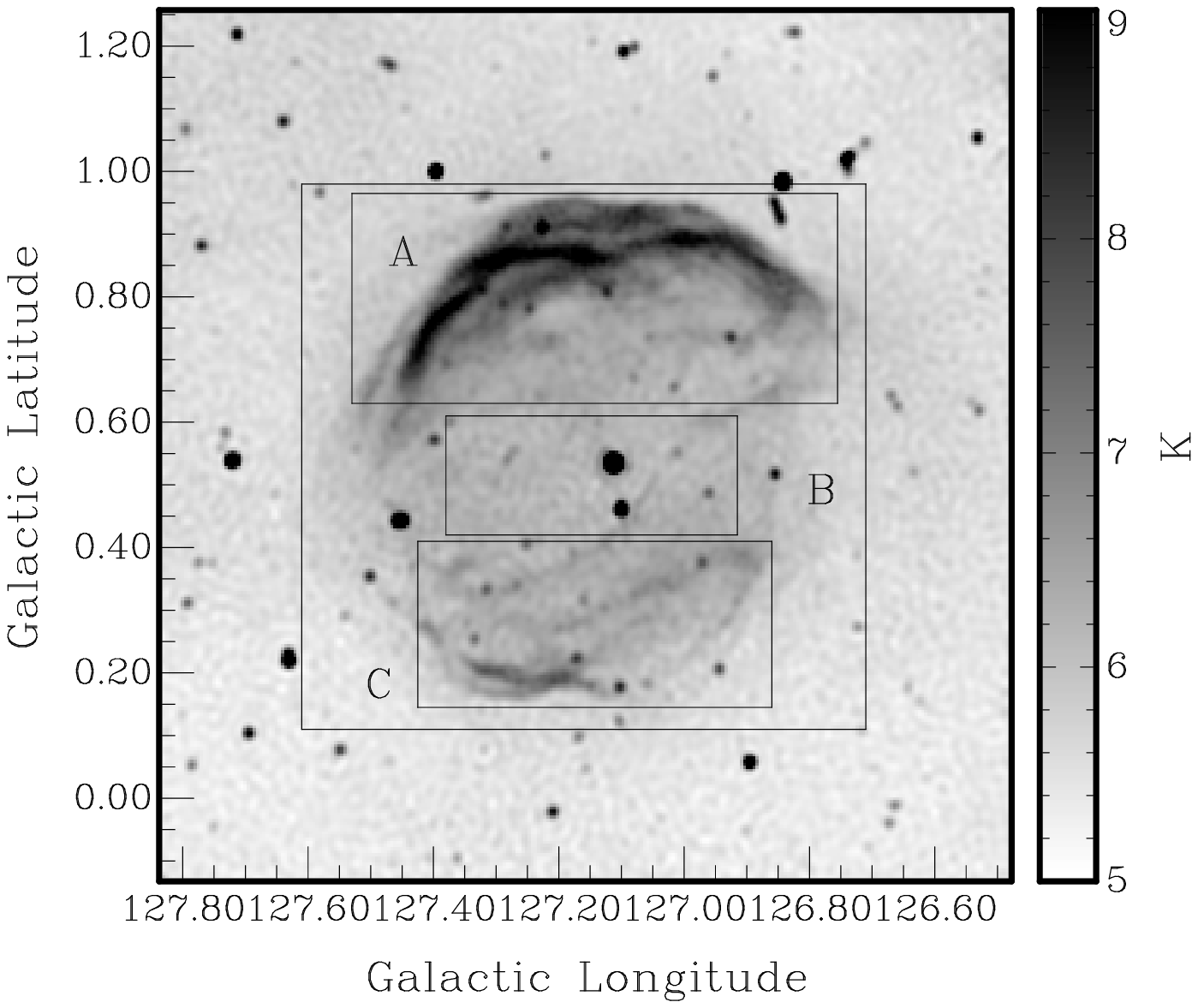}}
\put(10,-60){\includegraphics{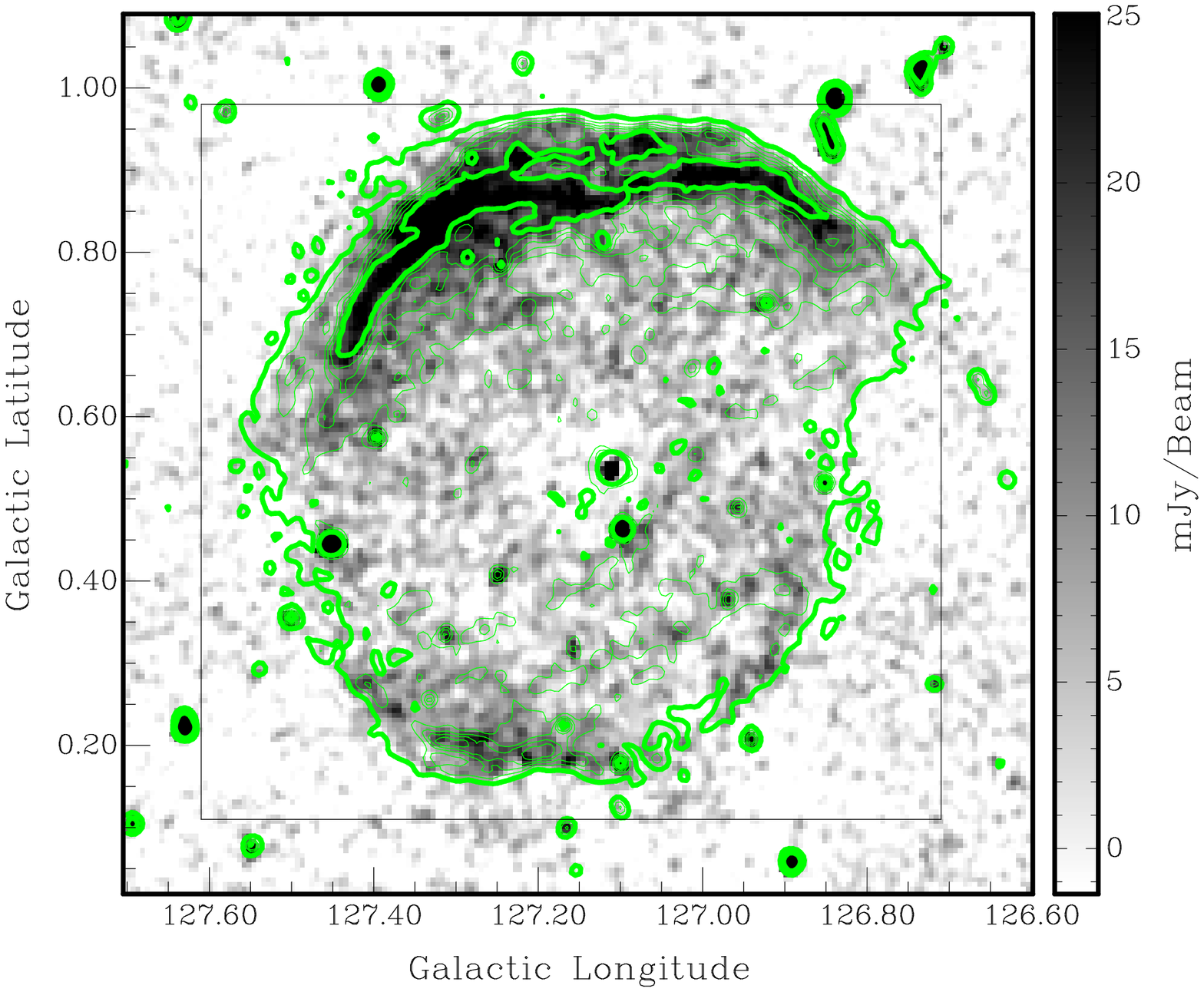}}
\put(260,-60){\includegraphics{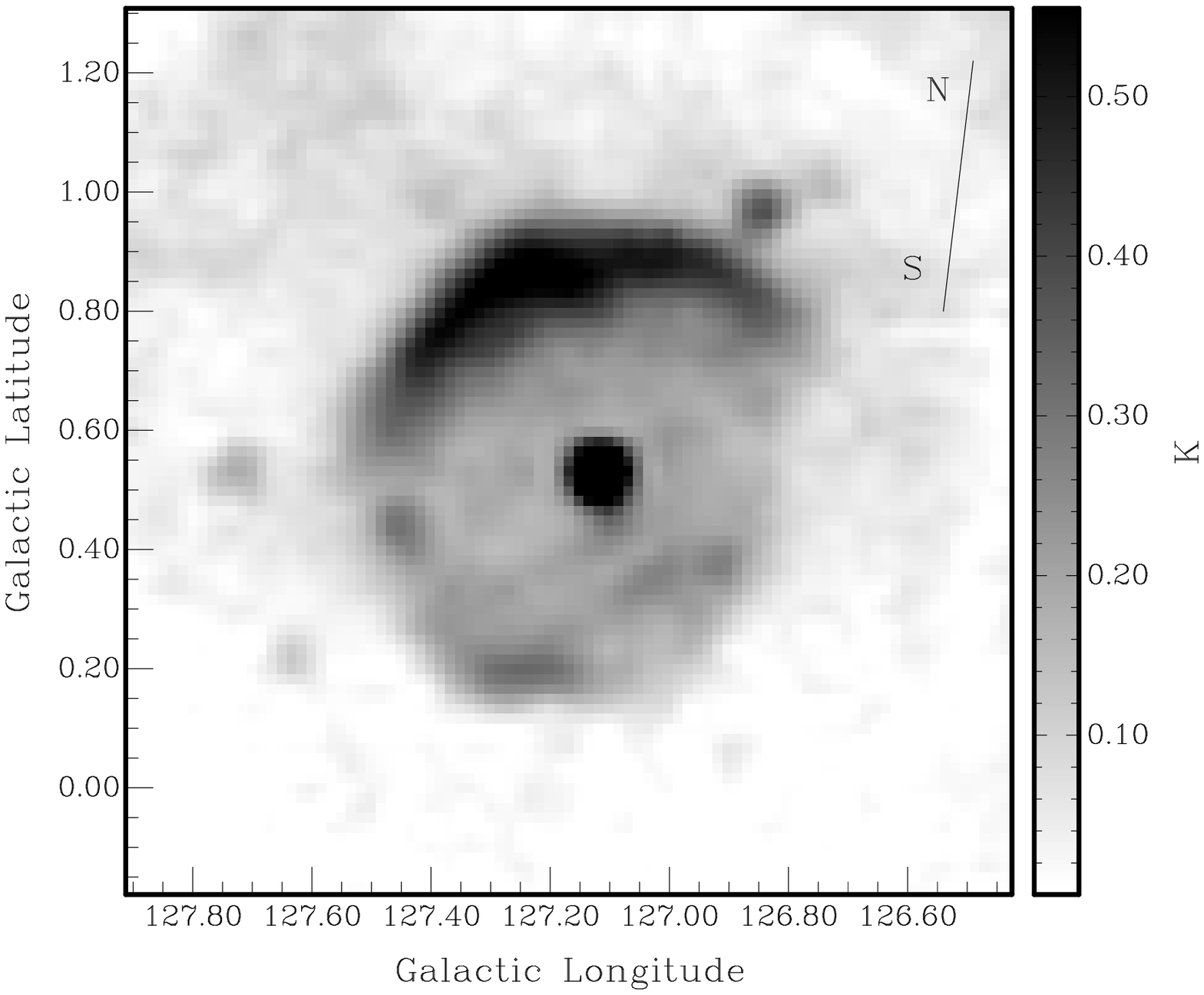}}
\end{picture}
\caption[xx]{The first row shows the CGPS maps at 
408 MHz (left) and 1420 MHz (right). The second row shows at left the WENSS image at 
327 MHz (grey scale) with contours from the 1420 MHz map (contours at 6, 6.5, 6.9, 7.2, 7.5, 8 K).  The second row shows at right the 
2695 MHz Effelsberg map. The box used for whole SNR T-T plots 
is shown in the upper left. The 3 boxes, labeled with letters and 
used for SNR sub-areas T-T plots, are shown in the upper right.
The direction of North (N) and South (S) is marked on the lower right image.}
\end{figure*}

\section{Results}
\subsection{Continuum Emission from G127.1+0.5}

The CGPS continuum images at 408 MHz and 1420 MHz are shown in the upper left and right
panels of Fig. 1. 
The lower left shows the WENSS (the Westerbork North Sky Survey, beam size  
54 $^{\prime \prime}$$\times $ 54 $^{\prime \prime}$ cosec($\delta$),  Rengelink et al., 1997) 
map at 327 MHz (grey scale) with contours from the 1420 MHz map. 
The 2695 MHz Effelsberg map is reproduced in the lower right for reference
(F\"urst et al., 1990). The Effelsberg map has a resolution of 4.3$^{\prime}$ and a brightness temperature sensitivity of 
50 mKT$_{B}$.

The 408 MHz image for G127.1+0.5 is similar to the 408 MHz image of Joncas et al.(1989). 
The southern and northern shell structures are clearly seen. The radius of northern shell is about 24$^{\prime}$, that of the southern is about 17$^{\prime}$.
The 1420 MHz map here 
shows much better detail of the fine structures in G127.1+0.5 than any previous image. 
A bright filament structure appears in the northern shell of the SNR, and a 
weak multi-filamentary structure appears in the southern part of the SNR.  
The 408 MHz image of G127.1+0.5 is consistent with a lower resolution version of the 
1420 MHz image, except for the intensity of the central source.
Fig. 1 shows that main features of G127.1+0.5 in the 327 MHz WENSS map are consistent with the CGPS 408 MHz and 1420 MHz maps. 

\begin{table}
\begin{center}
\caption{List of the 6 brightest compact sources and their integrated flux densities} 
\setlength{\tabcolsep}{1mm} 
\begin{tabular}{cccccc} 
\hline \hline
 No.&GLON& GLAT & S$_{408MHz}$&S$_{1420MHz}$& Sp. Index\\ 
\hline &deg &deg &mJy&mJy&$\alpha$\\ 
\hline 
\hline 
1&  126.840&0.980 &356 $\pm$56&187$\pm$34&0.52 (0.25 to 0.79)\\
2&  127.451&0.445 &264 $\pm$13& 91$\pm$ 4&0.86 (0.81 to 0.91)\\ 
3$^{a}$&  127.096&0.464 &164 $\pm$12& 62$\pm$ 3&0.78 (0.71 to 0.85)\\
4$^{b}$&  127.110&0.538 &123 $\pm$11&448$\pm$14&-1.04(-1.11to-0.96)\\
5&  127.272&0.529 & 39 $\pm$ 4&  7$\pm$ 1&1.42 (1.26 to 1.62)\\
6&  127.585&0.972 & 32 $\pm$ 3& 10$\pm$ 1&0.94 (0.84 to 1.06)\\
\hline
\hline
\end{tabular}
\end{center}
$^{a}$ named 0124+627 previously;
$^{b}$ named 0125+628 previously.
\end{table}

\begin{table*}
\begin{center}
\caption{Integrated flux densities of the compact source 0125+628}
\setlength{\tabcolsep}{1mm}
\begin{tabular}{cccccc}
\hline \hline
 S$_{408MHz}$&S$_{608MHz}$&S$_{1420MHz}$&S$_{2695MHz}$&S$_{5.0GHz}$&S$_{8.7GHz}$\\
\hline mJy&mJy&mJy&mJy&mJy&mJy\\
\hline
123$\pm$11$^{a}$&210$\pm$8$^{b}$&448$\pm$14$^{a}$&508$\pm$35$^{c}$(38*) &480$\pm$30$^{c}$(23*)&482$\pm$30$^{c}$(15*)\\
\hline
\end{tabular}
\end{center}
*flux density from nearby compact source 0124+627 which is included in present flux density of 0125-628 (see text). $^{a}$ this paper; $^{b}$Pauls et al. (1982); $^{c}$ Salter et al.(1978)
\end{table*}

\begin{table}
\begin{center}
\caption{408-1420 MHz T-T plot spectral indices
with and without Compact Sources(CS)}
\setlength{\tabcolsep}{1mm}
\begin{tabular}{ccc}
\hline
\hline
 Area  &$\alpha$ &  $\alpha$   \\
\hline
  &including CS &  CS removed \\
\hline
 A& $0.49\pm$0.01& 0.49$\pm0.01$\\
 B& $0.50\pm$0.17& 0.7$\pm$0.3 (manual fit)\\
 C& $0.46\pm$0.06& 0.46$\pm0.06$\\
\hline
All G127.1+0.5&0.47$\pm$0.01&0.46$\pm$0.01\\
\hline
 \hline
\end{tabular}
\end{center}
\end{table}

\begin{table}
\begin{center}
\caption{Integrated flux densities and spectral indices of G127.1+0.5 with and without compact sources(CS)}
\setlength{\tabcolsep}{1mm}
\begin{tabular}{cccc}
\hline
\hline
Source&S$_{408MHz}$ & S$_{1420MHz}$& $\alpha$\\
 &Jy&Jy &\\
\hline
G127.1+0.5& 17.1$\pm$1.7&10.0$\pm$0.8&0.43$\pm$0.10\\
G127.1+0.5 $\&$ CS&18.1$\pm$1.8&10.8$\pm$0.9&0.41$\pm$0.10\\
\hline
\hline
\end{tabular}
\end{center}
\end{table}

\begin{table}
\begin{center}
\caption{Integrated Flux Densities (FD) of G127.1+0.5}
\setlength{\tabcolsep}{1mm}
\begin{tabular}{cccc}
\hline
\hline
Freq. &HPBW&FD & references for FD\\
MHz   &arcmin  & Jy            & \\
\hline
\hline
 178 & 23$\times$19 & 37.0 $\pm$7.0 & 1978, Salter et al. \\
 408 &3.5$\times$3.9& 17.9 $\pm$2.0* &1989, Joncas et al.\\
 408 &3.4 $\times$3.8 & 17.1 $\pm$1.7* &  this paper\\
 865 & 14.5$\times$14.5&14.9 $\pm$ 0.8 &  2003, Reich et al.\\
 1420& 1$\times$1.1 & 10.1 $\pm$ 0.8* &  1989, Joncas et al.\\
 1420& 1$\times$1.1 & 10.0 $\pm$ 0.8* &  this paper\\
 2695 & 4.4$\times$4.4& 8.4 $\pm$ 0.6 &  1984, F\"urst et al.\\
 2695 & 4.3$\times$4.3& 8.3 $\pm$ 0.8 & see text\\
 4850 & 2.6$\times$2.6& 6.2$\pm$ 0.4  & 1984, F\"urst et al. \\
\hline
\hline
\end{tabular} 
\end{center}
* means CS within the SNR have been subtracted.
\end{table}

\subsection{T-T Plot Spectral Indices}
Bright compact sources affect the measured integrated flux densities for G127.1+0.5 and its measured spectral index $\alpha$. Thus we correct for the effects of compact sources. 
Table 1 lists properties of the 6 brightest compact sources which are detected
within G127.1+0.5 at both 408 MHz and 1420 MHz.   

First we discuss spectral indices between 408 MHz and 1420 MHz based on 
the T-T plot method.
The principle of the T-T plot method 
is that spectral indices (T$_{\nu}$=T$_{o}$$\nu$$^{-\beta}$) are calculated from a fit of a linear relation to the T$_{1}$-T$_{2}$ values of all pixels within a given map region.  T$_{1}$ is the brightness temperature of a map pixel at one frequency and T$_{2}$ is for the second frequency. The higher resolution image has been smoothed to the lower resolution for the T-T plot comparison. The brightness temperature spectral index $\beta$ is derived from the slope of the line. The error in spectral index is derived from the uncertainly in slope of the line.  The flux density spectral index $\alpha$ (S$_{\nu}$$\propto$$\nu$$^{-\alpha}$) is related to $\beta$ by $\beta$=$\alpha$+2. 
Spectral index refers to flux density spectral index $\alpha$ in this paper unless specifically noted otherwise.  

For the T-T plot analysis, first a single region for the whole SNR is used, 
as shown in Fig. 1. This region yields the T-T plots shown in Fig. 2. 
Two cases are considered: using all pixels including compact sources; 
and excluding compact sources.
Previous similar studies (Tian and Leahy, 2005a,2005b,2006; Leahy and Tian, 2005) have shown the 
second case produces better results compared to subtracting compact sources. 
The compact sources are usually bright compared to the SNR emission. 
Since the compact sources have generally a steeper spectrum than the SNR, 
they are seen in the T-T plot as steeper lines of points. 
But in case of G127.1+0.5, 0125+628 has a 
spectrum between 408 MHz and 1420 MHz which is much flatter than the SNR's. 
This shows up in the left panel of Fig. 2 as a flatter line of points.
Removing regions of pixels including 
compact sources from the analysis produces the right plot of Fig. 2. 

\begin{figure*}
\vspace{43mm}
\begin{picture}(60,80)
\put(-105,320){\includegraphics{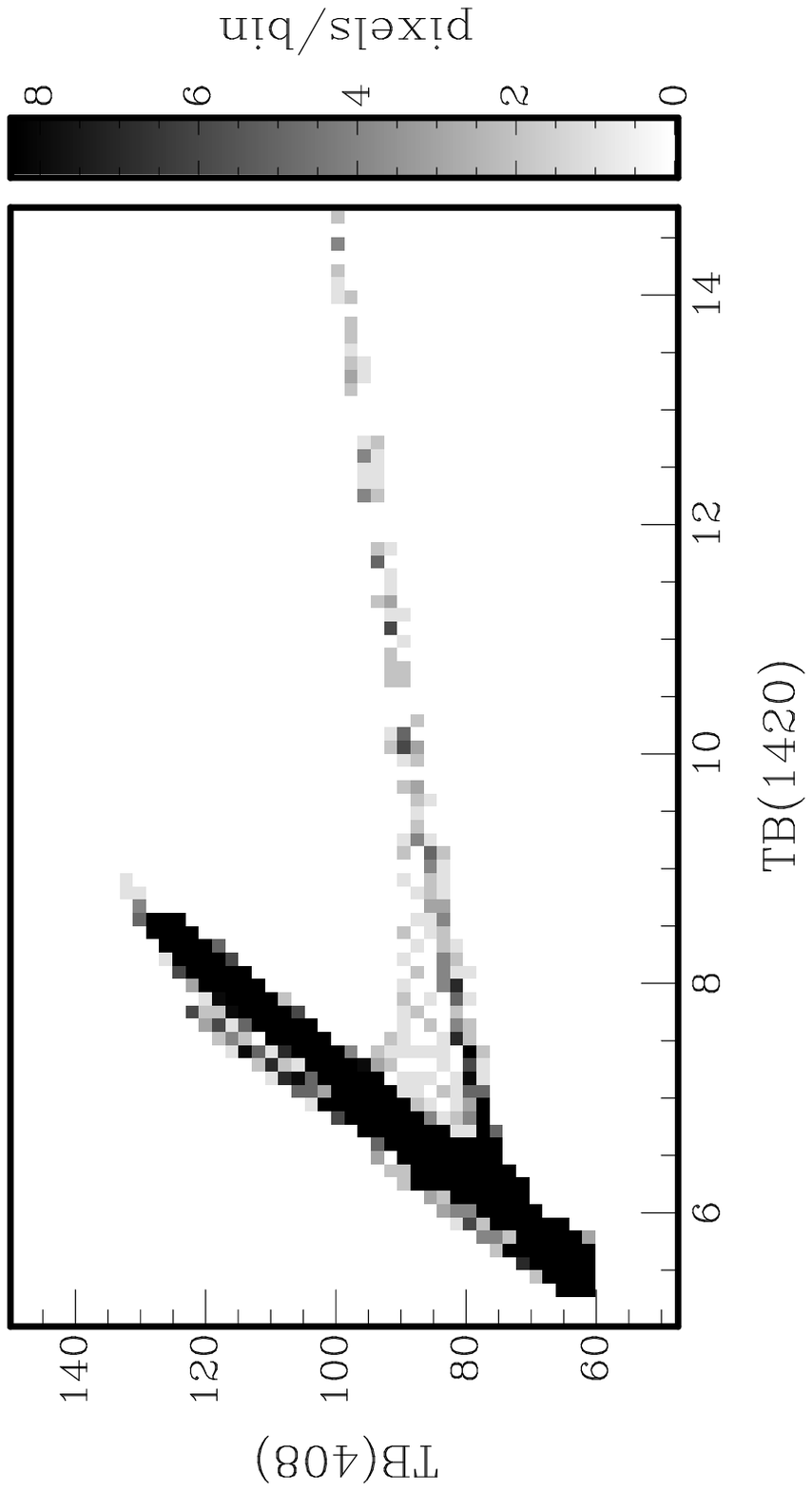}} 
\put(300,-40){\includegraphics{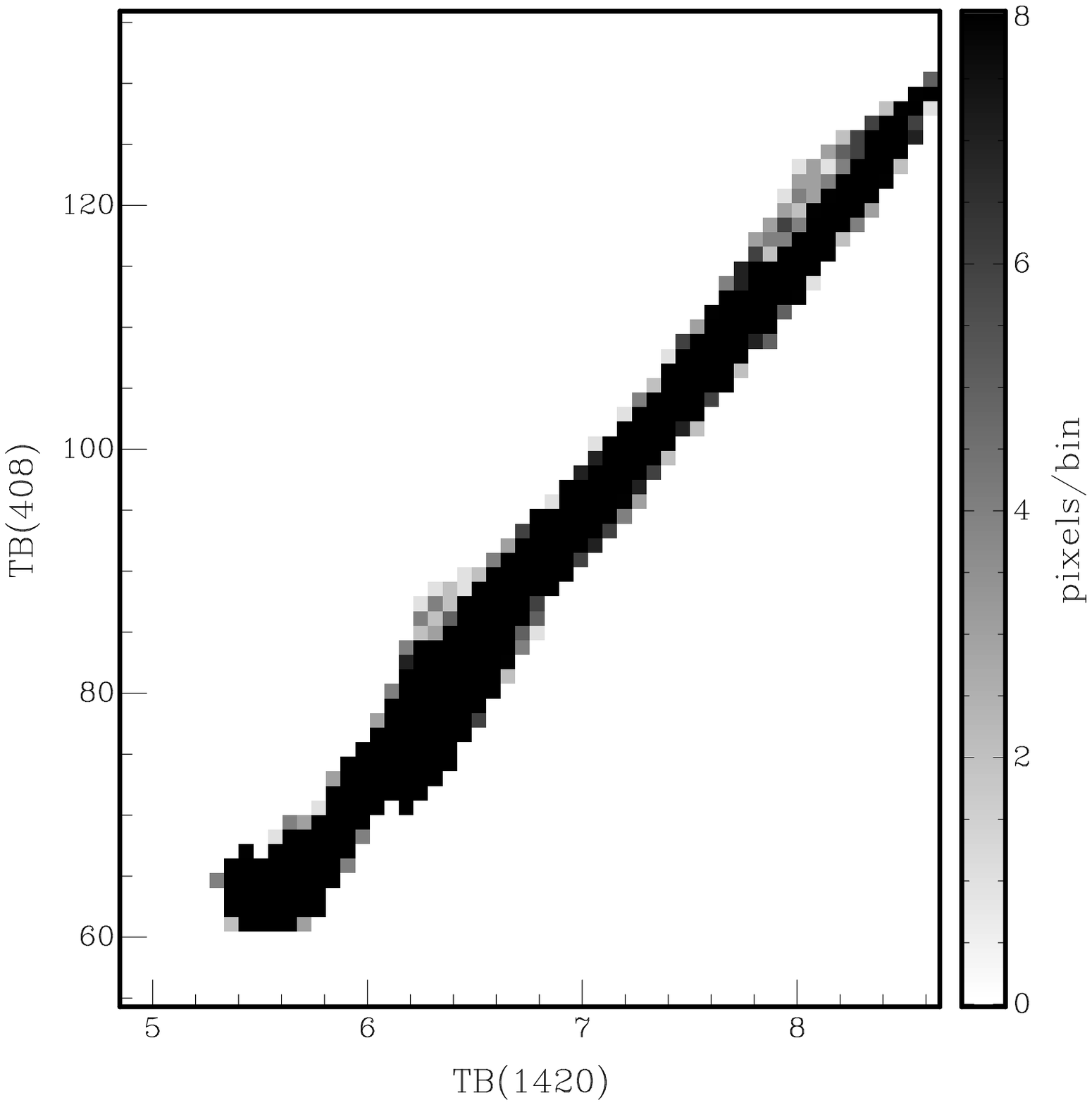}}
\end{picture}
\caption[xx]{Whole SNR 408-1420 MHz T-T plots.  The left plot includes compact sources 
($\alpha$=0.47$\pm$0.01); the right plot has compact sources removed from the analysis 
($\alpha$=0.46$\pm$0.01).}
\end{figure*}

Next, three smaller areas (labeled A to C in the upper right panel of
Fig. 1) are selected to search for possible spatial variations in spectral index.  
Table 3 lists the results for the two cases of analysis: 
including compact sources and removing compact sources. 
There are no significant compact sources removed from either southern and northern shell regions (areas A and B). The influence of compact sources' flux density on the spectral index calculation is strong for the central area B. However, since the diffuse SNR emission doesn't have much 
brightness variation the uncertainty in the T-T plot derived spectral index is large. 
From now on we discuss spectral indices derived with compact sources removed, unless specified otherwise. There is no evidence at 1$\sigma$ for spatial variations in spectral index within 
G127.1+0.5.

\begin{figure}
\vspace{55mm}
\begin{picture}(0,80)
\put(-20,-20){\includegraphics{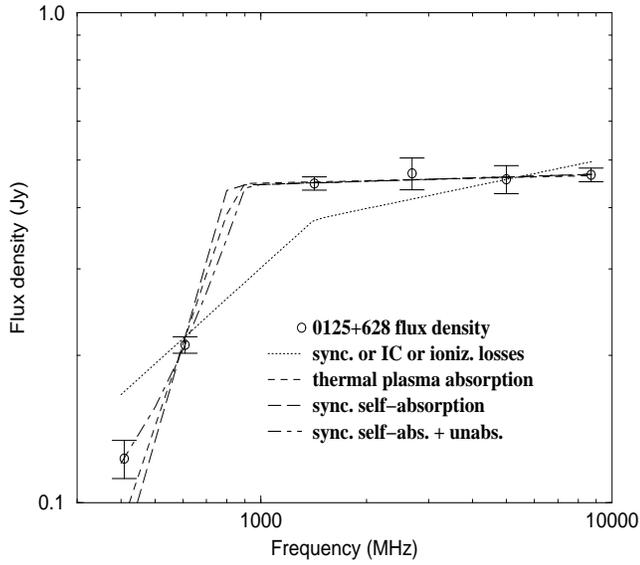}}
\end{picture}
\caption[xx]{Radio spectrum of 0125+628 and model fits (see text).}
\end{figure}

\begin{figure}
\vspace{55mm}
\begin{picture}(0,80)
\put(-20,-20){\includegraphics{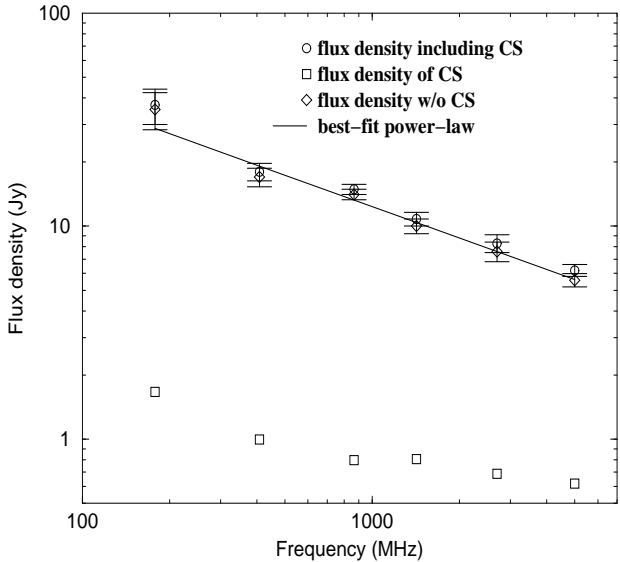}}
\end{picture}
\caption[xx]{Radio spectrum of G127.1+0.5. The best-fit spectral index is 0.49, with 
$\chi^2$=3.9, for a fit to 6 frequencies in range 178-4850 MHz.}
\end{figure}

\begin{figure*}
\vspace{95mm}
\begin{picture}(60,60)
\put(-10,-55){\includegraphics{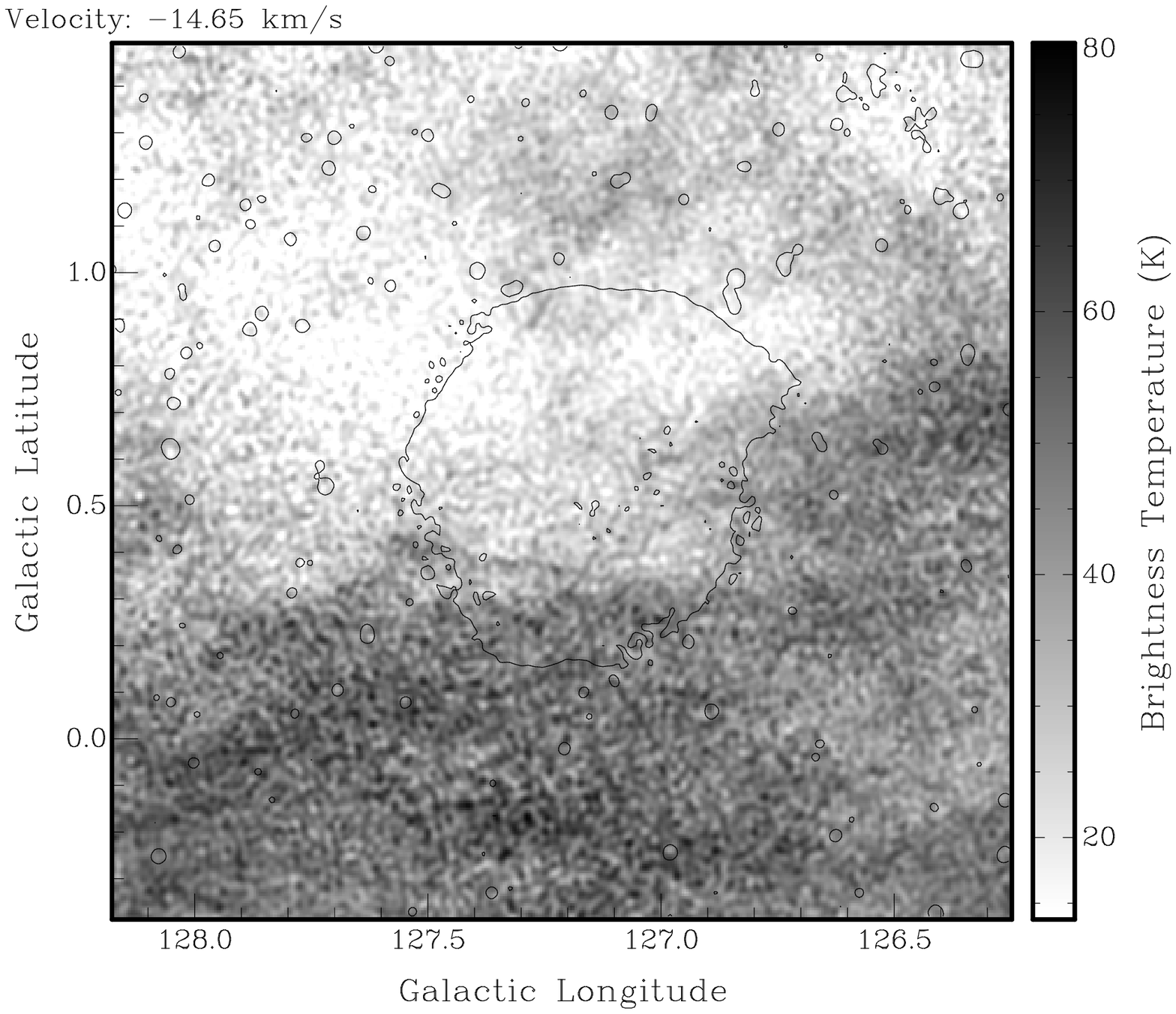}}
\put(163,-55){\includegraphics{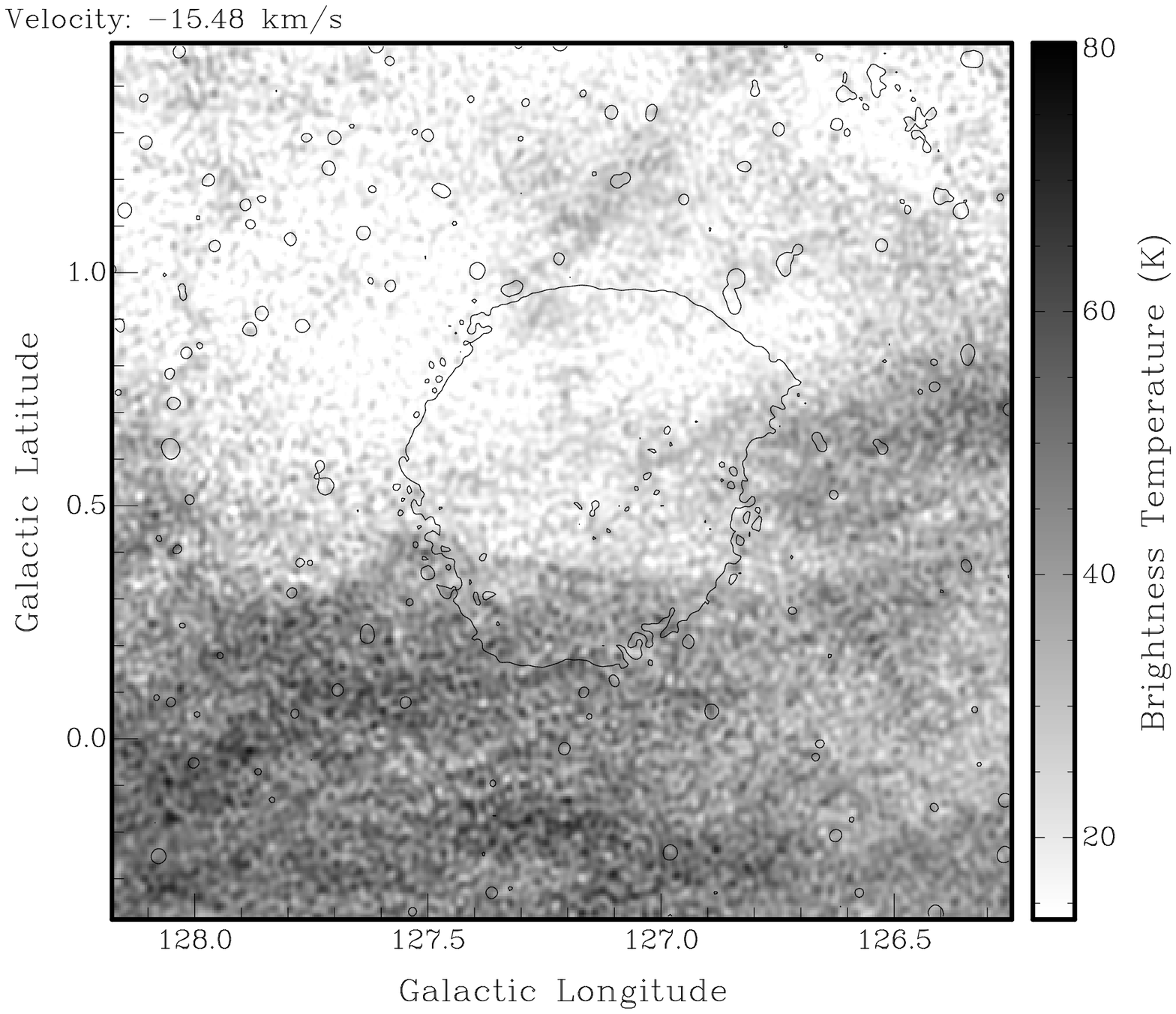}}
\put(335,-55){\includegraphics{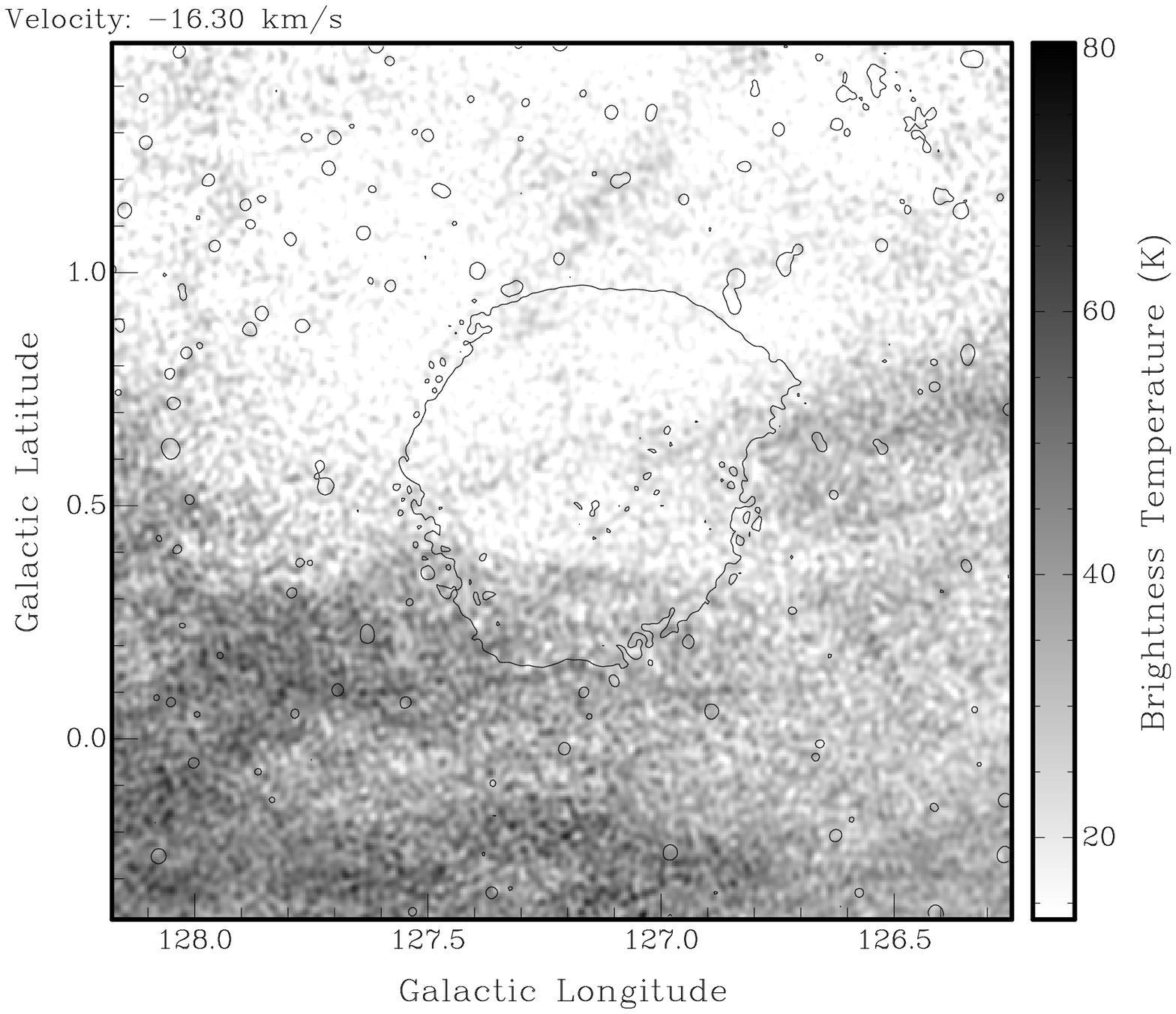}}
\put(-10,115){\includegraphics{G127-Hi-12.eps}}
\put(163,115){\includegraphics{G127-Hi-13.eps}}
\put(335,115){\includegraphics{G127-Hi-13-1.eps}}
\end{picture}
\caption[xx]{HI emission in the field centered on G127.1+0.5 from -12 to -16 km$/$s. The radial velocity of each map
is indicated at its top left corner. The outline of G127.1+0.5 in 1420 MHz continuum emission
is indicated by the contour at 6 KT$_{B}$.}
\end{figure*}

\subsection{Central Compact Sources within the SNR}

Before discussing G127.1+0.5, we consider the central sources since the flux densities of the
SNR must be corrected for the flux densities of all compact sources including the central sources.
Flux densities for the two central compact sources 0124+627 and 0125+628 were determined from
the 408 MHz and 1420 MHz maps. 
0124+627 is a steep spectrum compact source with spectral index of 0.78 between 408 MHz and 1420 MHz. 
We use this spectral index and its flux density at 1420 MHz to estimate its flux density in other frequencies.  
0125+628's low-frequency turnover radio spectrum 
is confirmed based on data at six frequencies (Table 2, Fig. 3).
The current analysis is improved comparing with previous work (Joncas et al., 1989; Salter et al., 1978), since we have subtracted the contributions from 0124+627 to 0125+628's flux densities, which were included in the previous studies. 

Different model spectra for 0125+628 are considered here (e.g. Longair, 1981; 
Leahy $\&$ Roger, 1998). 
If the spectrum has a turnover due to electron energy losses, it could be due to synchrotron or
inverse Compton losses at high energy or due to ionization losses at low energy. In the former case
the electron power-law injection spectral index, $\gamma$, is related to the low frequency 
radio spectral index, $\alpha_{low}$ by $\alpha_{low}=(\gamma-1)/2$ and the radio spectral index
at high frequency is increased by 0.5. In the latter case, the electron power-law 
injection spectral index, $\gamma$, is related to the high frequency 
radio spectral index, $\alpha_{hi}$ by $\alpha_{hi}=(\gamma-1)/2$ and the radio spectral index
at low frequency is decreased by 0.5. 
The best fit model for both cases is shown in Fig. 3. It has $\alpha_{low}$=-0.65, 
$\alpha_{hi}$=-0.15 with break frequency 1400 MHz and $\chi^2$=49. It is inconsistent with the
data as it doesn't give a sharp enough turnover. 
Absorption mechanisms give a sharper turnover.
Absorption by a dense thermal plasma in front of the radio source gives an exponential cutoff in
frequency which is too sharp, but one can have absorption by a thermal plasma intermixed with
the radiating electrons, which gives a power-law cutoff where the low frequency radio spectral
index is decreased by 2.1 with respect to the high frequency index. This model fits the data
with $\alpha_{low}$=-2.26, 
$\alpha_{hi}$=-0.16 with break frequency 860 MHz and $\chi^2$=8.3. 
Synchrotron self-absorption gives a low frequency spectral index of -2.5, and also fits the data
with best fit parameters:
$\alpha_{hi}$=-0.02 with break frequency 810 MHz and $\chi^2$=16.2. 
A better fitting model is one for a source with synchrotron self-absorption but with some part of
the source not subject to self-absorption. This can easily occur if the source has regions with
high and low density of synchrotron emitting electrons. 
This model fits the data very well ($\chi^2$=0.2) with 
$\alpha_{hi}$=-0.02 with break frequency 900 MHz for the high density region, and relative 
contribution of the low density region (with $\alpha$ fixed equal to $\alpha_{hi}$) of 20$\%$. 
In summary, a simple model with synchrotron
self-absorption fits the spectrum of 0125+628. This model is used to calculate the contribution
of 0125+628 to the integrated flux density of G127.1+0.5 at frequencies where 0125+68 has
not been directly measured.

\subsection{Integrated Flux Densities and Spectral Indices}

Integrated flux densities of G127.1+0.5 with diffuse background subtracted were derived at 
408 MHz and 1420 MHz.
The resulting 408 MHz to 1420 MHz spectral index, using flux densities without compact sources, 
is 0.43$\pm$0.10.  
Table 4 lists the flux densities at 408 and 1420 MHz and spectral indices with and without the compact sources within G127.1+0.5. 
Compact sources contribute about 6$\%$ at 408 MHz and 8$\%$ at 1420 MHz, 
and generally have no significant effect on spectral index except for area B. 
The whole SNR spectral index derived from integrated flux densities is
consistent with the whole SNR spectral index (0.46$\pm$0.01) derived
by the T-T plot method.

Published integrated flux densities and errors for the SNR 
at other frequencies are given in Table 5. 
For 408 and 1420 MHz the values in Table 5 already have compact source flux density
removed. F\"urst et al (1984) gave a flux density based on the 2965 MHz Effelsberg image with resolution 4.4 $\times$ 4.4 arcmin, but we obtain a new value from the Effelsberg 2695 MHz image with a little higher resolution 4.3 $\times$ 4.3 arcmin 8.3$\pm$0.8 Jy and use this instead.
We fit the resulting flux density values with a power-law to obtain the spectral index.  Figure 4 shows the corrected flux densities and the best-fit power-law with $\alpha$=0.49, 90$\%$ uncertainty
range 0.42 to 0.56.  
     
\subsection{HI Emission}

We have searched the CGPS HI images for features in the HI which might relate to the morphology of G127.1+0.5.
There is emission along the edge of northern and southern shells of G127.1+0.5 in the velocity 
range -12 to -16 km/s, and only in this range. 
Fig. 5 shows maps of HI emission in the six velocity channels in this range. 
One contour of 1420 MHz continuum emission is shown to indicate the outline of G127.1+0.5.
The emission along the edge of G127.1+0.5 can alternately be described as 
a depression in the HI emission.
This HI association with G127.1+0.5 is similar to other accepted HI associations 
with supernova remnants and with similar HI velocity ranges ($\sim$10-20 km/s), 
e.g. for the Cygnus Loop (Leahy, 2003) and DA530 (Landecker et al., 1999). 

\section{Discussion}

\subsection{The Distance and Age of G127.1+0.5}

There are no previous reliable distance estimates for G127.1+0.5.  
Joncas et al. (1989) suggested a range of 2 - 5 kpc. 
Previous observations didn't provide any compelling evidence to associate G127.1+0.5 with 
either the compact sources within its area (Spinrad et al. 1979; Pauls et al. 1982, Goss $\&$ van Gorkom 1984) or with the superposed cluster NGC 559 (Pauls et al. 1982). 
Fu\"rst et al. (1984) inspected images from the Maryland 21 cm HI Line survey (Westerhout, 1972) 
and reported a hole between velocities -70 to -90 km$\/$s centered on G127.1+0.5. Our high resolution HI images do not confirm this hole: only in velocity range -12 to -16 km$/$s, 
is there emission coincident with G127.1+0.5 (Fig. 5).
  
The HI radial velocity for G127.1+0.5 of -14 km$/$s gives a distance estimate 
based on the galactic rotation curve. We take R$_{0}$=8.5 kpc, V$_{0}$=220 km s$^{-1}$. 
If G127.1+0.5 is within $\sim$2 kpc, then the circular velocity at the SNR is also 220 km s$^{-1}$,
which gives a distance of 1.15 kpc. 
If we make the extreme assumption that G127.1+0.5 is far away and that 
the circular velocity at the SNR is 250 km s$^{-1}$ we find an upper limit to the distance
of 2.9 kpc. 
The open cluster NGC 559 located on the face of the SNR has a distance 0.9 - 1.3 kpc 
(Blair et al. 1980, Xilouris et al. 1993), consistent with our estimate of G127.1+0.5's distance. 
The association of NGC 559 with G127.1+0.5 is also supported by detection of optical emission 
in the area of G127.1+0.5 (Xilouris et al. 1993) which generally indicates a close distance to us. 
 
At 1.15 kpc distance, G127.1+0.5 has a mean radius of r=8 pc. 
Summing the HI channel maps from -12 to 16 km/s, taking the difference between the rim and 
center of G127.1+0.5, and converting to column density, we find an excess of 
$\simeq10^{20}$cm$^{-2}$ along the rim. 
This yields an estimate of the local density $n_0=10^{20}$cm$^{-2}/(2r) \simeq 2 $cm$^{-3}$. 
Applying a Sedov model (e.g. Cox, 1972), for a typical explosion energy of 
E=0.5$\times10^{51}$ erg ($\epsilon_{0}$=E/$(0.75\times 10^{51}erg)$=2/3), yields an age of 
$\simeq$3$\times$10$^{4}$ yr.  
However, the Sedov age is large enough that G127.1+0.5 should have entered the dense shell phase.
For $n_0=2cm^{-3}$ the shock radius and the age when time the SNR enters the dense shell phase are 15.8 pc
and $2.3\times10^4$ yr, so G127.1+0.5 should be just past entering the dense shell phase.
The Sedov estimate for shocked gas temperature just interior to the shell at the time of 
the start of the dense shell phase is still valid: $7\times 10^5$K. 
In the cooling phase, the material which
is recently shocked cools rapidly to add mass to the dense shell. 
The dense shell emits mostly in the ultraviolet, which will not be observable
through the column density to the SNR. This is $\simeq 2\times 10^{21}cm^{-2}$, 
derived from either the 
extinction to NGC 559 or the HI channel maps for velocities $<-14$km/s. Similarly the low
luminosity soft x-ray emission from the SNR interior would not be visible, consistent with
its lack of detection in the Rosat All-Sky Survey data.

\subsection{Radio Spectra of G127.1+0.5 and 0125+628}

Fig. 4 shows the radio spectrum of G127.1+0.5. 
Excluding compact source flux densities, the spectrum is consistent ($\chi^{2}$=3.9) 
with a power-law with $\alpha$=0.49, 90$\%$ confidence range of 0.42 to 0.56.
The compact sources do not strongly affect the SNR's spectrum. 
The spectral index from integrated flux densities agrees with both the T-T plot spectral index 
(0.46$\pm$0.01) and the 408-1420 MHz integrated flux density spectral index (0.43$\pm$0.10). 

The central compact sources 0125+628 and 0124+627 have been previously studied because 
they are located near the center of the SNR. 
Optically 0125+628 was associated with an elliptical galaxy by Kirshner and Chevalier (1978). 
Pauls et al. (1982) confirm that 0125+628 is a background radio source due to its large column
density. 
Kaplan et al. (2004) find a hard spectrum x-ray source associated with 0125+628 consistent with
0125+628 being an AGN, but no x-ray emission from 0124+627.
Spitler and Spangler (2005) observe 0125+628 with the VLBA and find no excess scattering associated
with G127.1+0.5. 

For 0124+627, we find no evidence of deviation from a power-law spectrum comparing our 408 and
1420 MHz data to the 608 MHz value of  Pauls et al. (1982).
For 0125+628 we obtain a significantly improved spectrum (Fig. 3) by correcting published values for
confusion with 0124+627.
The strong spectral break cannot be reproduced by 
synchrotron, inverse Compton, or ionization loss models.
Synchrotron self-absorption or thermal plasma absorption (if mixed with the emitting plasma) 
can give the observed break, with the thermal plasma absorption model somewhat better fitting due to
its smaller spectral break. However since the synchrotron self-absorption takes place only in high
density emitting regions, one might expect to have some fraction of the source at low density and 
free of self-absorption. This last case provides the best fit to the observed spectrum and yields an
unabsorbed fraction of $\simeq 20\%$. Since 0125+628 is known to be an AGN, one would expect 
synchrotron self-absorption similar to what is seen for a number of other AGN. Thus we prefer
the last model as a description of the observed spectrum.

\section{Conclusion}  

We present new maps of G127.1+0.5 at 408 MHz and 1420 MHz and study the radio spectrum, corrected
for point source flux densities, using different methods.  All results are consistent with the 
best determined value from the T-T plot method of 0.46$\pm$0.01. We find no evidence for spectral 
index variation across G127.1+0.5. A probable association of HI features with G127.1+0.5
yields a distance of 1.15 kpc for G127.1+0.5, consistent with association with the open cluster 
NGC 559. The HI deficit associated with the SNR gives a local density of $\simeq 2 cm^{-3}$, which
yields a dense shell transition radius of 7.9 pc, roughly equal to the observed radius. So G127.1+0.5 should be approximately at the age where it enters the dense shell phase, and have an age
of $2-3\times 10^4$yr.
  
\begin{acknowledgements}
We acknowledge support from the Natural Sciences and Engineering Research Council of Canada.  
The DRAO is operated as a national facility by the National Research Council of Canada.  
The Canadian Galactic Plane Survey is a Canadian project with international partners. 
\end{acknowledgements}


\begin{thebibliography}{}
\bibitem[1980]{Blaet80}Blair, W.P., Kirshner, R.P., Gull, T.R., Sawyer, D.L., Parker, R.A.R., 1980, ApJ, 242, 592
\bibitem[1972]{Coxet72}Cox, D., 1972, ApJ, 178, 159
\bibitem[1984]{Furet84}F\"urst, E., Reich, W. and Steube, R., 1984, A$\&$A, 133, 11
\bibitem[1990]{Furet90}F\"urst, E., Reich, W., Reich, P. and Reif, K., 1990, A$\&$AS, 85, 691
\bibitem[1984]{Goset84}Goss, W.M.$\&$ van Gorkom, J.H., 1984, A$\&$A, 5, 425
\bibitem[1982]{Haset82}Haslam, C.G.T., Salter, C.J., Stoffel, H. and Wilson, W.W., 1982, A\&AS, 47, 1
\bibitem[2000]{Higet00}Higgs, L.A. and Tapping, K.F., 2000, AJ, 120, 2471
\bibitem[1989]{Jonet89}Joncas, G., Roger, R.S. and Dewdney, P.E., 1989, A$\&$A, 219, 303
\bibitem[2004]{Kapet04}Kaplan, D.L., Frail, D.A., Gaensler, B.M. et al. 2004, ApJS, 153, 269
\bibitem[2004]{kir78}Kirshner, R.P. and Chevalier, R.A, 1978, Nature, 276, 480 
\bibitem[1999]{Lanet99}Landecker, T.L., Routledge, D., Reynolds, S.P., et al. 1999, ApJ, 527, 866
\bibitem[1981]{Long81}Longair, M.S., 1981, High Energy Astrophysics (Cambridge University Press)
\bibitem[2003]{Leaet98}Leahy, D.A. and Roger, R.S., 1998, ApJ, 505, 784
\bibitem[2003]{Leaet03}Leahy, D.A., 2003, AJ, 586, 224
\bibitem[2005]{Leaet05}Leahy, D.A. and Tian, W.W., 2005, A$\&$A, 440, 929 
\bibitem[1988]{Milet88}Milne, D.K., 1988 in: Supernova Remnants and the Interstellar Medium, ed. Roger, R.S. and Landecker, T.L., Cambridge: University Press, 351 . 
\bibitem[1977]{Pauet77}Pauls, T., 1977, A$\&$A, 59, L13
\bibitem[1982]{Pauet82}Pauls, T., van Gorkom, J. H., Goss, W.M., Shaver, P.A., Dickey, J.M., Kulkarni, S., 1982, A$\&$A, 112, 120 
\bibitem[1990]{Reiet90}Reich, W., Reich, P. and F\"urst, E., 1990, A$\&$AS, 83, 539
\bibitem[1997]{Reiet97}Reich, W., Reich, P. and F\"urst, E., 1997, A$\&$AS, 126, 413
\bibitem[2003]{Reiet03}Reich, W., Zhang, X. and F\"urst, E., 2003, A$\&$A, 408, 961
\bibitem[1997]{Renet97}Rengelink, R.B., Tang, Y., de Bruyn, A.G., Miley, G.K., Bremer, M.N., R\"ottgering, H.J.A., Bremer, M.A.R., 1997, A$\&$AS 124, 259
\bibitem[1978]{Salet78}Salter, C.J., Pauls, T., Haslam, C.G.T., 1978, A$\&$A, 66, 77
\bibitem[2005]{Spiet05}Spitler L.G. and Spangler S.R., 2005, ApJ., 632, 932
\bibitem[1979]{Spiet79}Spinrad, H., Stauffer, J., Harlan, E., 1979, PASP, 91, 619
\bibitem[2003]{Tayet03}Taylor, A.R., Gibson, S.J.,  Peracaula, M. et al., 2003, AJ, 125, 3145
\bibitem[2005a]{Tiaet05}Tian, W.W. and Leahy, D.A., 2005a, A$\&$A, 436, 187
\bibitem[2005b]{Tibet05}Tian, W.W. and Leahy, D.A., 2005b, A$\&$A, accepted, Astro-ph: 0508276
\bibitem[2005b]{Tibet05}Tian, W.W. and Leahy, D.A., 2006, ChJAA, accepted, Astro-ph: 0511320
\bibitem[1972]{Weset72}Westerhout, G., 1972, Maryland-Greenbank Galactic 21 cm Survey
\bibitem[1993]{Xilet93}Xilouris, K.M., Papamastorakis, J., Paleologou, E.V., Andredakis, Y., Haerendel, G., 1993, A$\&$A, 270, 393
 \end{thebibliography}
\end{document}